\documentclass[12pt,a4paper]{article}
\usepackage{a4wide}
\usepackage{latexsym}
\usepackage{epsf}
\usepackage{amssymb}

\makeatletter
\@addtoreset{equation}{section}
\makeatother

 \def\unit{\hbox to 3.3pt{\hskip1.3pt \vrule height 7pt width .4pt \hskip.7pt
\vrule height 7.85pt width .4pt \kern-2.4pt
\hrulefill \kern-3pt
\raise 4pt\hbox{\char'40}}}

\def\half{{\textstyle {1 \over 2}}}
\def\makeatletter{\catcode`\@=11}
\makeatletter
\def\mathbox#1{\hbox{$\m@th#1$}}%
\def\math@ccstyles#1#2#3#4#5#6#7{{\leavevmode
      \setbox0\mathbox{#6#7}%
      \setbox2\mathbox{#4#5}%
      \dimen@ #3%
      \baselineskip\z@\lineskiplimit#1\lineskip\z@
      \vbox{\ialign{##\crcr
             \hfil \kern #2\box2 \hfil\crcr
             \noalign{\kern\dimen@}%
             \hfil\box0\hfil\crcr}}}}
\def\mathaccstyles{\math@ccstyles\maxdimen}
\def\maththroughstyles{\math@ccstyles{-\maxdimen}}
\def\unitmatrixDT%
 {\maththroughstyles{.45\ht0}\z@\displaystyle {\mathchar"006C}\displaystyle 1}

\begin{document}

\begin{flushright}
\small
UG/00-02\\
DAMPT-2000-33\\
MRI-PHY/P20000308\\
{\bf hep-th/0003221}\\
March 24, 2000
\normalsize
\end{flushright}

\begin{center}


\vspace{.7cm}

{\Large {\bf T-duality and Actions for Non-BPS D-branes}}

\vspace{.7cm}

 E.~A.~Bergshoeff${}^1$, M.~de Roo${}^1$, T.~C.~de Wit${}^1$,
      E.~Eyras${}^2$ and S.~Panda${}^3$

\vskip 0.4truecm

 ${}^1$Institute for Theoretical Physics\\
   Nijenborgh 4, 9747 AG Groningen\\
     The Netherlands\\
\vskip 0.4truecm
 ${}^2$DAMTP\\
   University of Cambridge\\
   Wilberforce Road\\
   Cambridge CB3 9EW, UK\\
\vskip 0.4truecm
 ${}^3$Mehta Research Institute of Mathematics and Mathematical Physics\\
   Chhatnag Road, Jhoosi\\
   Allahabad 211019, India\\

\vskip 1.0cm


{\bf Abstract}

\end{center}

\begin{quotation}

\small

We employ T-duality to restrict the tachyon dependence of
 effective actions for non-BPS D-branes.
 For the Born-Infeld part the criteria of T-duality and supersymmetry
 are satisfied by a simple extension
 of the D-brane Born-Infeld action.

\end{quotation}

\newpage

\pagestyle{plain}

\setcounter{section}{0}
\section{Introduction}

Dp-branes \cite{Po} have played a crucial rule in the understanding of
 the relations between different string theories. They are stable extended
 objects, which preserve half of the maximal supersymmetry. They are
 known in terms of explicit solutions to low-energy supergravity equations,
 and their effective actions and its symmetries are by now well
 understood \cite{Po2}.

D-branes also give rise to other stable and
 unstable objects, such as brane-antibrane configurations, and
 non-BPS D-branes  \cite{Sen1,Sen2,Sen3}.
 These objects and their descendants
 could potentially play an equally important role, since
 they extend the relations between string theories to a different
 domain\footnote{For reviews, see \cite{Sen4,Lerda, S1}}.
 Non-BPS branes in  the Type II theories are unstable,
 and can decay to the stable
 D-branes. Much work has already been done on the classification of these
 objects \cite{Wit1,Ho1},
 and this has clarified their relation to D-branes, as well
 as the structure of the hierarchy of D-branes themselves.

A proposal for an effective action for non-BPS D-branes in the Type II
 theories was given by Sen \cite{Sen5}.
 In this action the instability is due to the presence of a tachyon. The
 tachyon dependence must be such that condensation to the D-brane is
 possible, which puts severe restrictions on the tachyon dependence
 of this action. The general structure is that a non-BPS Dp-brane in
 the Type IIA (IIB) theory, will condense to a BPS
 D(p-1)-brane in the same IIA (IIB) theory. The non-BPS Dp-branes
 in IIA (IIB)  are related to BPS D(p+1)-brane and antibrane configurations
 in IIA (IIB) by condensation of the complex tachyon living on this
 brane-antibrane pair. This interrelationship between BPS and non-BPS branes
 implies that the T-duality map between Dp-branes in IIA (IIB) to
 D(p$\pm$1)-branes in IIB (IIA) must also hold for the non-BPS branes.
 This T-duality map gives further information about the tachyon dependence
 of the effective action of the non-BPS brane. It is our aim to investigate
 the T-duality properties of non-BPS branes in the Type II theories.

In this introduction we will give a short overview of the effective action
 for non-BPS branes, mainly following \cite{Sen5} for the Born-Infeld
 contribution, and \cite{Billo,Ken} for the Wess-Zumino term. Then we will
 discuss aspects of T-duality, first for the Born-Infeld term in Section 2,
 then for the Wess-Zumino term in Section 3. In  Section 4 we
 speculate about the analogue of tachyon condensation in non-BPS
 D-branes in the context of $D=11$ M-branes.

The Born-Infeld term in the action for  a non-BPS Dp-brane in a nontrivial
 background should be of the following
 form:
\begin{equation}
\label{SBI}
   S^{(p)}_{{\rm BI}} = -\int d^{p+1}\sigma e^{-\phi} \sqrt{|{\rm det}\, G_{ij}|}
        \, f(T,\partial T,..)\,,
\end{equation}
where
\begin{equation}
   G_{ij} = g_{ij} + {\cal F}_{ij}\,.
\end{equation}
Here $g$ is the metric induced by the supersymmetric line-element, and
 ${\cal F}$ involves the Born-Infeld vector
 $F_{ij}=2\partial_{[i}V_{j]}$ and the Neveu-Schwarz
 $B$-superfield. The function $f$ contains the dependence on the tachyon and
 its derivatives, and may also depend on other worldvolume and background
 fields.

 In this paper we discuss the conjecture  that (\ref{SBI})
  is of the form
 \begin{equation}
 \label{nonBPS}
    S_{\rm BI}^{(p)} =  -\int d^{p+1}\sigma \sqrt{ |{\rm det}\, ({G}_{ij} +
      \partial_iT\partial_jT)|}\,g(T)\,.
 \end{equation}
 The argument which we will advance in support of this conjecture is that
  (\ref{nonBPS}) agrees with T-duality and supersymmetry. Further arguments,
  based on a calculation of $S$-matrix elements, have been discussed
  by Garousi \cite{Gar}. The action (\ref{nonBPS}) is a special case
  of (\ref{SBI}), and can be rewritten in that form by expanding the
  root. Our result implies that all terms in the expansion
  of (\ref{nonBPS})
  satisfy the requirement of T-duality.
  Of course, this does not prove the conjecture.
  A proof would require an extension of \cite{Gar} to higher order terms
  in $\partial T$. In this respect it would be interesting to calculate
  the explicit form of the $(\partial T)^4$ contribution to the
  4-tachyon amplitude\footnote{We
  thank A.~Tseytlin for a discussion on this point.}.
  Our discussion of T-duality is useful independently of
  the validity of the conjecture. The formula (\ref{nonBPS}) generates a
  series of terms that agree with T-duality and supersymmetry. Even though
  in the complete answer the coefficients of these terms might differ from
  those that follow from (\ref{nonBPS}), our method enables us to identity
  the structure of the interactions that are allowed to appear in
  (\ref{SBI}).

The Wess-Zumino term for a single non-BPS D-brane
 takes on the form:
\begin{equation}
\label{SWZ}
   S^{(p)}_{\rm WZ} = \int d^{p+1}\sigma\,\, C\wedge dT\wedge e^{\cal F}\,.
\end{equation}
The superfield $C$ contains the Ramond-Ramond fields.
 This generalizes the D-brane Wess-Zumino action given by \cite{GHT},
 to which we refer for the notation. The leading form in $C$ is
 a $p$-form. The kink solution for the tachyon is expected to give a
 $\delta$-function from the $dT$-contribution and thus to produce the
 standard Wess-Zumino term for the resulting D(p-1)-brane.

The terms (\ref{SBI})
 and (\ref{SWZ}) are separately invariant
 under worldvolume reparametrizations and target space
 (super-)reparametrizations.
 It is assumed that the tachyon is a scalar under worldvolume
 reparametrizations, and that the function $f$ depends only on
 invariant combinations of worldvolume and background fields.
 However, the sum of (\ref{SBI}) and (\ref{SWZ}) is not $\kappa$-symmetric
 as it would be for D-branes \cite{Aga,Ced,BT}. The relation between
 the non-BPS Dp-brane and the BPS D(p-1)-brane then arises as follows.
 The tachyonic kink-solution effectively reduces the dimension of the
 worldvolume by one. Because the tachyon is then constant almost everywhere,
 terms with derivatives of tachyons vanish. The remainder of the action
 should then be the standard effective action of a D(p-1)-brane.
 The fermionic $\kappa$-symmetry, which is absent for the non-BPS
 brane, is restored by the tachyon condensation.
 In the resulting action the full set of background fields, as well as
 invariance under all (super-)reparametrizations, are still present.

In a flat background this can be made more explicit (see \cite{Aga}).
 Then we have\footnote{Worldvolume indices are denoted by $i,j=0,\ldots,p$,
 target space indices by $\mu,\nu=0,\ldots,D-1$.}:
\begin{eqnarray}
\label{gij}
   g_{ij}&=&\eta_{\mu\nu}\Pi^\mu_i\Pi^\nu_j\,;\qquad
   \Pi^\mu_i = \partial_iX^\mu - \bar\theta\Gamma^\mu\partial_i\theta\,,
   \\
\label{Fij}
   {\cal F}_{ij}&=&\partial_iV_j-\partial_jV_i
    -\left\{\bar\theta\Gamma_{11}\Gamma_\mu\partial_i\theta
     \left(\partial_jX^\mu - \half\bar\theta\Gamma^\mu\partial_j\theta\right)
         -(i\leftrightarrow j)\right\}\,, \\
\label{Cij}
    C_{i_1\ldots i_p} &=& \partial_{[i_1} X^{\mu_1}
  \cdots \partial_{i_{p-1}}X^{\mu_{p-1}}\,\bar\theta{\cal P}_{(p)}
   \Gamma_{\mu_1\ldots\mu_{p-1}}
     \partial_{i_p]}\theta + \ldots\,.
\end{eqnarray}
In (\ref{gij}-\ref{Cij}) we present the IIA case. The Majorana
 spinor $\theta$ can be expanded as $\theta=\theta_L+\theta_R$. To obtain the
 IIB case we should replace $\Gamma_{11}$ by $\sigma_3$ and write
 $\theta$ as a doublet $(\theta_{1R}\ \theta_{2R})$. In the
 expression for the Ramond-Ramond field
 ${\cal P}_{(p)}$ equals $\Gamma_{11}$ for $p=4k+1$ and $\unitmatrixDT$
 for $p=4k+3$. In the IIB case $p$ is even, and we must have
 ${\cal P}_{(p)}$ equal to $i\sigma_2$ for $p=4k$, and equal to
 $\sigma_1$ for $p=4k+2$. In (\ref{Cij}) we have not written
 higher-order contributions in the fermions, which are required for
 supersymmetry.

\section{T-duality and the Born-Infeld term}

Before coming to non-BPS D-branes, let us briefly recall how T-duality
 works for D-branes \cite{BdR}.
 In this example we work in a flat background, but we keep the worldvolume
 fermions to identify later
 the possible couplings between fermions and tachyons.
 For a  Dp-brane we have the following Born-Infeld term
\begin{equation}
   {\cal L}_{\rm BI} = -\sqrt{ |{\rm det}\, {G}^{(p)}_{10\,ij}|}\,,
\end{equation}
with ${G} = {g} + {\cal F}$. We will reduce a IIA Dp-brane and a IIB
 D(p+1)-brane to a nine dimensional Dp-brane. T-duality amounts to the fact
 that the resulting worldvolume actions
  should be the same in the two cases. The reduction
 of the fermions in the IIA case is as follows \cite{Cvet}:
\begin{equation}
   \theta_R \to \left(\theta_1 \atop 0\right)\,,\quad
   \theta_L \to \left( 0 \atop \theta_2\right)\,.
\end{equation}
The $\Gamma$-matrices reduce as
\begin{equation}
   \Gamma^\mu \to
    \left(
    \begin{array}{cc} 0&\gamma^\mu\\ \gamma^\mu & 0 \end{array}
    \right)
  \,,\quad(\mu = 0,\ldots,8),\quad
   \Gamma^9 \to
    \left(
    \begin{array}{cc} 0& \unitmatrixDT \\ -\unitmatrixDT & 0 \end{array}
    \right)
    \,,\quad
   \Gamma^{11} =
    \left(
    \begin{array}{cc} \unitmatrixDT&0\\ 0 &-\unitmatrixDT \end{array}
    \right) \,.
\end{equation}
The reduction is over a transverse direction and the corresponding
 coordinate $X^9$ is written as a worldvolume scalar $S$. The
 result is
\begin{eqnarray}
   {G}^{(p)}_{10\,ij} &\to& G^{(p)}_{9\,ij} - \partial_iS\partial_jS
    + 2\bar\theta_2\partial_i\theta_2\partial_jS
    - 2\bar\theta_1\partial_j\theta_1\partial_iS
   + 2\bar\theta_2\partial_i\theta_2\bar\theta_1\partial_j\theta_1\,,
\end{eqnarray}
with
\begin{eqnarray}
    G^{(p)}_{9\,ij} &=&
   g_{ij} + F_{ij} -2\bar\theta_2\gamma_\mu\partial_i\theta_2\partial_jX^\mu
                   -2\bar\theta_1\gamma_\mu\partial_j\theta_1\partial_iX^\mu
  \nonumber\\&&
   + \bar\theta_2\gamma_\mu\partial_i\theta_2
     \bar\theta_2\gamma^\mu\partial_j\theta_2
   + \bar\theta_1\gamma_\mu\partial_i\theta_1
     \bar\theta_1\gamma^\mu\partial_j\theta_1
   + 2 \bar\theta_2\gamma_\mu\partial_i\theta_2
     \bar\theta_1\gamma^\mu\partial_j\theta_1 \,.
\end{eqnarray}
In the IIB case we reduce a D(p+1)-brane over a worldvolume direction.
 We gauge-fix the corresponding coordinate $X^9$ equal to a worldvolume
 coordinate $\sigma$, and the corresponding component of the Born-Infeld
 vector becomes a worldvolume scalar $S$. The fermions now reduce as follows:
\begin{equation}
   \theta_{1R} \to \left(\theta_1 \atop 0\right)\,,\quad
   \theta_{2R} \to \left(\theta_2 \atop 0\right)\,,
\end{equation}
and we obtain the following:
\begin{equation}
   G^{(p+1)}_{10} \to
 \left(\begin{array}{cc}
      G^{(p)}_{9\,ij} - 2\bar\theta_2\partial_i\theta_2
     \bar\theta_1\partial_j\theta_1   &
     \partial_iS -2\bar\theta_2\partial_i\theta_2\\
     -\partial_jS -2\bar\theta_1\partial_j\theta_1& -1
        \end{array}\right)\,.
\end{equation}
Finally we have to prove that the determinants of the two nine-dimensional
 expressions are the same. This can be shown by using the identity
\begin{equation}
\label{matrix-id}
   {\rm det}  \left(\begin{array}{cc}
                     A & B \\
                     C & D
               \end{array}\right)
  = {\rm det} \left(\begin{array}{cc}
            A - BD^{-1}C  & B \\
                 0        & D
          \end{array}\right)
     \left(\begin{array}{cc}
              \unitmatrixDT & 0 \\
             D^{-1} C & \unitmatrixDT
           \end{array}\right)
   = {\rm det}\, (  A - BD^{-1}C) {\rm det}\, D\,.
\end{equation}

To identify possible couplings between tachyons and the other worldvolume
 fields in the case of the non-BPS D-brane, we consider as
 the starting point the Lagrangian in (\ref{nonBPS}):
\begin{equation}
\label{LnonBPS}
   {\cal L}_{\rm BI}^{(p)} = -\sqrt{ |{\rm det}\, ({G}^{(p)}_{10\,ij} +
     \partial_iT\partial_jT)|}\,g(T)\,.
\end{equation}
If we assume that $T$ becomes independent of the compact direction in which
 the T-duality transformation is performed, then the result of the
 calculation we performed above will be the same, with the replacement
\begin{equation}
   G^{(p)}_{9\,ij} \to G^{(p)}_{9\,ij} + \partial_iT\partial_jT\,.
\end{equation}
The equality between the determinants  still holds, and T-duality
 will be preserved. If in addition we assume that $T$ is inert under
 supersymmetry, (\ref{LnonBPS}) is also supersymmetric.

If we expand (\ref{LnonBPS}) in the tachyon field, we reobtain an
 action of the form  (\ref{SBI}), with a set of explicit couplings
 of the tachyon to worldvolume fields, and, in a general background,
 target space fields.
 Let us write explicitly the leading terms, including
 expressions quadratic in the fermions and in $\partial T$,
 which result from this expansion for the case of a IIA flat background:
\begin{eqnarray}
\label{nonBPSexp}
   {\cal L}^{(p)}_{\rm BI} &=& g(T)\,\sqrt{ |{\rm det}\,
       G_{ij} |}\times
  \bigg\{1+
   {1\over 2}G^{ji}(-2\bar\theta_L\Gamma_\mu\partial_i\theta_L\partial_jX^\mu
          -2\bar\theta_R\Gamma_\mu\partial_j\theta_R\partial_iX^\mu
          +\partial_i T\partial_jT) \nonumber\\
  &&-{1\over 2} G^{ki}(-2\bar\theta_L\Gamma_\mu\partial_i\theta_L\partial_lX^\mu
          -2\bar\theta_R\Gamma_\mu\partial_l\theta_R\partial_iX^\mu)
      G^{lm}\partial_mT\partial_kT\nonumber\\
  &&+{1\over 4} G^{ji}(-2\bar\theta_L\Gamma_\mu\partial_i\theta_L\partial_jX^\mu
          -2\bar\theta_R\Gamma_\mu\partial_j\theta_R\partial_iX^\mu)
       G^{kl}\partial_lT\partial_kT
      + \ldots\bigg\}\,,
\end{eqnarray}
where $G_{ij}= \eta_{\mu\nu}\partial_iX^\mu\partial_jX^\nu +F_{ij}$,
 and $G^{ij}$ is its inverse. These couplings satisfy the requirement of
 T-duality. They are not supersymmetric by themselves, since
 we have expanded the supersymmetric combinations (\ref{gij}, \ref{Fij}).
 However, supersymmetry can be restored by
 the addition of quartic fermion terms, which follow
 from (\ref{gij}, \ref{Fij}). The expression (\ref{nonBPSexp})
 was written for the IIA case, for IIB replace $\theta_L\to \theta_{2R}$,
 $\theta_R\to \theta_{1R}$.

In the literature there has been some effort to include couplings between
 worldvolume fermions and tachyons in the function $f$ \cite{Klus}. This work
 suggests a coupling of the form
\begin{equation}
   \partial_i\bar\theta\partial_j\theta G^{ij}\,,
\end{equation}
multiplying a function of $T$. Although this form is indeed nonzero in the
 IIA theory, it does not have a counterpart in the IIB theory because
 of the different chirality structure.
 Therefore it does not satisfy T-duality, and should not
 appear in the non-BPS brane action.

 From (\ref{LnonBPS}) it is clear that the $(\partial T)^2$ terms couple only
 to the symmetric part of $G^{(p)\,-1}_{10}$. Nevertheless, in the expansion
 (\ref{nonBPSexp})  the terms mixing $(\partial T)^2$ with the
 fermionic contributions do couple to a nonvanishing NS-NS background field.

Note that the expansion of the determinant in (\ref{LnonBPS}) gives a series
 of couplings between tachyons and other fields, with fixed
 relative coefficients. Supersymmetry and T-duality are not sufficient
 to fix these coefficients; any couplings of tachyons
 with (\ref{gij}, \ref{Fij}) would satisfy these two requirements.
 Also, the potential $g(T)$ in (\ref{LnonBPS}) is
 not restricted by our arguments. Recently, remarkable progress has been
 made in constructing $g(T)$ from open string field
 theory \cite{Sen6,Sen7,Tay,Berk}.

\section{T-duality and Wess-Zumino terms}

For the Wess-Zumino terms we will restrict ourselves to contributions
 quadratic in the fermions, and again to a flat background.
 Let us first consider the case of a non-BPS p-brane ($p$ odd) in the IIA
 theory. The Lagrangian in
 (\ref{SWZ}) then takes on the explicit form
\begin{equation}
\label{LWZIIA}
  {\cal L}^{(p)}_{\rm WZ} =
  \epsilon^{i_1\ldots i_{p+1}} \sum_{k=0}^{(p-1)/2} \,\,a_{p,k}
  C_{i_1\ldots i_{p-2k}}(F^k)_{i_{p-2k+1}\ldots i_p} \partial_{i_{p+1}}T\,,
\end{equation}
where $C$ is given by (\ref{Cij}). The coefficients $a_{p,k}$ will be
fixed by T-duality (see below).
Writing out the Majorana spinor
 $\theta$ in terms of chiral components, we find, after a partial integration
\begin{equation}
   C_{i_1\ldots i_{p-2k}}=
 2\partial_{[i_1}X^{\mu_1}\cdots \partial_{i_{p-2k-1}}X^{\mu_{p-2k-1}}\,
   \bar\theta_L\Gamma_{\mu_1\ldots \mu_{p-2k-1}}
   \partial_{i_{p-2k}]}\theta_R\,.
\end{equation}
Upon reduction to $D=9$ the total result becomes
\begin{eqnarray}
  {\cal L}^{(p)}_{\rm WZ}&\to&
  2\epsilon^{i_1\ldots i_{p+1}} \sum_{k=0}^{(p-1)/2} \,a_{p,k}\,\bigg\{
    \tilde C_{i_1\ldots i_{p-2k}} (F^k)_{i_{p-2k+1}\ldots i_p}
     \partial_{i_{p+1}}T
   \nonumber\\
  &&\ + (p-2k-1)
 \tilde C_{i_1\ldots i_{p-2k-1}}\partial_{i_{p-2k}}S
   (F^k)_{i_{p-2k+1}\ldots i_p}
     \partial_{i_{p+1}}T\bigg\} \,.
\end{eqnarray}
Here we have defined
\begin{equation}
  \tilde C_{i_1\ldots i_m}=
 \partial_{[i_1}X^{\mu_1}\cdots \partial_{i_{m-1}}X^{\mu_{m-1}}\,
   \bar\theta_2\gamma_{\mu_1\ldots \mu_{m-1}}
   \partial_{i_{m}]}\theta_1\,,
\end{equation}
which are the nine-dimensional RR-fields in a flat $D=9$ background.

Again we should compare with the result obtained by reducing the WZ-term of
 a IIB non-BPS $p+1$-brane to nine dimensions.
 The starting point is of the same form as (\ref{LWZIIA}), except that we
 will use coefficients $b_{p+1,k}$.
 Recall that the worldvolume
 scalar $S$ now comes from the world-volume vector. The final result is
\begin{eqnarray}
     {\cal L}^{(p+1)}_{\rm WZ}&\to&
 - 2\epsilon^{i_1\ldots i_{p+1}} \,s_{p-2k+1} b_{p+1,k}\bigg\{
    \sum_{k=0}^{(p-1)/2} \,
    (p-2k)\tilde C_{i_1\ldots i_{p-2k}} (F^k)_{i_{p-2k+1}\ldots i_p}
     \partial_{i_{p+1}}T
   \nonumber\\
  &&\ + \sum_{k=1}^{(p+1)/2}\,2k
 \tilde C_{i_1\ldots i_{p-2k+1}}
   (F^k)_{i_{p-2k+2}\ldots i_{p-1}}\partial_{i_p}S
     \partial_{i_{p+1}}T\bigg\} \,.
\end{eqnarray}
Here $s_m$ ($m$ even) is a sign which is $+1$ for $m=4l+2$, and $-1$ for
 $m=4l$.

T-duality determines the coefficients $a_{p,k}$ and $b_{p+1,k}$ up to
 an overall normalization.
 The result is
\begin{eqnarray}
  a_{p,k} &=& {(p-1)!\over 2^k\, k!\, (p-2k-1)!}\, a_{p,0}\,,
  \nonumber\\
  b_{p+1,k}&=& {(-1)^k p!\over 2^k\, k! \,(p-2k)!}\, b_{p+1,0}\,,
  \nonumber\\
  a_{p,0} &=& -s_{p+1}\, p\, b_{p+1,0}\,.
\end{eqnarray}
The normalization relative to the Born-Infeld term can be fixed
 by requiring that the Dp-brane action
 which arises for the kink solution is $\kappa$-symmetric.

The coefficients $a$ and $b$ in this section turn out to be the same
 as those that are required by T-duality of Dp-branes. We conclude therefore
 that also the Wess-Zumino terms of non-BPS branes satisfy T-duality.

\section{Discussion}

We have shown, assuming a particular form of the tachyon coupling,
that the worldvolume action for non-BPS D-branes proposed by Sen
satisfies the criteria of T-duality and supersymmetry.
The particular form (\ref{nonBPS}) is interesting by itself.
The same form has been suggested, from a different point of view,
in \cite{Gar}. It
 suggests
 a higher-dimensional structure for the non-BPS brane, a point which was
 remarked also by Ho\u rava \cite{Ho1}. It is therefore natural to discuss
non-BPS branes within the context of the eleven-dimensional M-theory.
In M-theory open M2-branes can end on M5-branes. The quantization of
the open M2-brane, in a certain low-energy limit, leads to a self-dual
tensor multiplet living on the worldvolume of the M5-brane.
An important difference with the ten-dimensional context is that
it is not
known what the field theory  describing a set of coinciding M5-branes
or a M5-$\overline{\rm M5}$ system should be.

On the side of classification, which depends mainly on the
 structure of the Wess-Zumino terms, a lot of work in $D=11$
 has already been done \cite{Lo1,Lo2}. In this context an interesting
relation between K-theory and the Killing isometry direction of
\cite{Killing} has been pointed out \cite{Vancea}.

As far as the dynamics is concerned the situation is more complicated.
The field-theoretical approach
 which is suitable for a tachyon field, resulting from open strings, is
 replaced by a more complicated structure involving strings on the
 worldvolume, which represents interactions with membranes.
 A number of interesting points about the issue of
 ``tachyonic string'' condensation have been raised by Yi
 \cite{Yi}. In this scenario the M5-$\overline{\rm M5}$ system decays into
a BPS M2-brane.
To achieve this Yi proposes a Higgs-mechanism for the non-selfdual tensor that
should arise from the two selfdual tensors living on the
M5-$\overline{\rm M5}$ system. An unattractive feature of this scenario
is that the source for the three transverse scalars that are needed to
describe the M2-brane remain unclear. It would be desirable if
a Higgs-mechanism could be constructed for a single self-dual tensor only.
The other remaining selfdual tensor exactly contains the three degrees
of freedom which are required to describe the three transverse scalars.

In order to see whether a Higgs mechanism for a single selfdual tensor
can be constructed it is convenient to use as a starting point
the following action:

\begin{equation}
{\cal L}_0 = -{\textstyle{1\over 24}}\epsilon^{ijklm}H_{0ij}H_{klm}
-{\textstyle{1\over 12}}H_{ijk}H^{ijk}\,,
\end{equation}
with $i=1,\ldots,5$.
This action is a gauge-fixed version of the action constructed in \cite{PST}.
Note that the Lagrangian is not Lorentz covariant but the equations of
motion are. A necessary condition for the existence of a Higgs mechanism
is that a massive extension of the Lagrangian ${\cal L}_0$ exists.
The only local mass term one can write down is given by
\footnote{One could consider
the Lagrangian for the sum of a selfdual and an anti-selfdual tensor
and then try to add a mass term for the diagonal combination only.
This would not change our discussion.}

\begin{equation}
{\cal L}_m = m^2 B^{\mu\nu} B_{\mu\nu}\,.
\end{equation}
However, we have checked that, although the equations of motion corresponding
to ${\cal L}_0$ and ${\cal L}_m$ are Lorentz covariant, the ones
corresponding to the combination ${\cal L} = {\cal L}_0 + {\cal L}_m$ violate
Lorentz symmetry. Therefore, within the context of a local field theory,
a Higgs mechanism for a single self-dual tensor seems not possible.
It will be interesting to see how such a mechanism is realized
in M-theory.

\section{Acknowledgements}

We would like to thank David Berman, Ben Craps, Jan de Boer and
 Robbert Dijkgraaf
 for useful discussions.  This work is supported by the European
 Commission TMR program ERBFMRX-CT96-0045, in which E.B.~and M.d.R.~are
 associated to the University of Utrecht. The work of T.C.d.W. is part of
 the research program of the ``Stichting voor Fundamenteel Onderzoek der
 Materie'' (FOM). E.E.\ is supported by the European Community program
 ``Human Potential'' under contract HPMF-CT-1999-00018 and partially
 by a PPARC grant PPA/G/S/1998/00613. S.P.\ thanks the Institute for
 Theoretical Physics in Groningen, where this work was initiated,
 for its warm hospitality.

\end{document}